\documentclass[12pt]{aastex}
\usepackage{emulateapj5}

\slugcomment{submitted to ApJ Letters, June 13, 2000; accepted July
17, 2000}

\begin{document}
\title{Multicolor Observations of a Planetary Transit of HD 209458}

\author{Saurabh Jha\altaffilmark{1}, David Charbonneau\altaffilmark{1,2}, Peter
M. Garnavich\altaffilmark{1,3}, Denis J. Sullivan\altaffilmark{4},
Tiri Sullivan\altaffilmark{4}, Timothy M. Brown\altaffilmark{2}, 
John L. Tonry\altaffilmark{5}}
\email{sjha@cfa.harvard.edu, dcharbonneau@cfa.harvard.edu}

\altaffiltext{1}{Harvard-Smithsonian Center for Astrophysics, 60 Garden
Street, Cambridge, MA 02138}
\altaffiltext{2}{High Altitude Observatory, National Center for Atmospheric
Research, P.O. Box 3000, Boulder, CO 80307-3000. NCAR 
is sponsored by the National Science Foundation.}
\altaffiltext{3}{Department of Physics, University of Notre Dame, 
225 Nieuwland Science Hall, Notre Dame, IN 46556}
\altaffiltext{4}{School of Chemical and Physical Sciences, Victoria University
of Wellington, P.O. Box 600, Wellington, New Zealand}
\altaffiltext{5}{Institute for Astronomy, University of Hawaii, 
Honolulu, HI 96822}

\begin{abstract}
We present $BVRIZ$ photometric observations of HD~209458 during the
transit by its planetary companion on UT 1999 November 15 with the
University of Hawaii 0.6m and 2.2m telescopes and the High Altitude
Observatory STARE telescope. The detailed shape of the transit curve
is predicted to vary with color due primarily to the color-dependent
limb-darkening of the star, but potentially due as well to the effect
of color-dependent opacity in the planetary atmosphere.  We model the
light curves and present refined values for the transit timing and
orbital period, useful for planning future observations of the
planetary transit. We also derive significantly improved measurements
of the planetary radius, $R_p = 1.55 \pm 0.10 \, R_{\rm Jup}$, stellar
radius, $R_s = 1.27 \pm 0.05 \, R_{\sun}$, and orbital inclination, $i =
85.9 \pm 0.5 \degr$. The derived planetary radius favors evolutionary
models in which the planet has a low albedo.
\end{abstract}


\section{Introduction}

The discovery of transits of the Sun-like star HD~209458 by an
orbiting, low-mass companion has definitively established the
existence of extrasolar gas-giant planets \citep{char00,henr00}.  The
identification of a transiting extrasolar planet system allows, for
the first time, exploration of the physical characteristics of a
planet outside our solar system.

The most interesting quantity that we can derive from the photometric
observations of the transit is the planetary radius, $R_{p}$.
Although HD~209458b is less massive than Jupiter, its radius is
significantly larger, a result of slowing of the planetary contraction
due to exposure to high stellar insolation early after the planet's
formation (Burrows et al. 2000). The precise value of $R_{p}$ is
dependent upon the planetary Bond albedo, $A$, a fundamental parameter
in the thermodynamics of the planetary atmosphere.  A precise
measurement of $R_{p}$ may allow an estimate of $A$ \citep{burr00},
though the published uncertainty in $R_{p}$ \citep{maze00} is too
great for a meaningful constraint. 

The photometric observations of the transit published to date have
been in the $R$ band (two full transits; \cite{char00}) and in a
Str\"omgren $b+y$ band (one half transit; \cite{henr00}).  In this
\emph{Letter}, we present observations in five wavelength bands of the
planetary transit on UT 1999 November 15. As we describe below, we
have observed color-dependent variations in the transit shape due to
the stellar limb-darkening.  We present a likelihood analysis of the
data that improves the estimates of the planetary radius and orbital
inclination.  More generally, we outline a method that may be used to
analyze future high-precision, multicolor data and exploit the
color-dependence in order to break the degeneracy shared between the
planetary and stellar parameters.

\section{Observations and Data Reduction}
\subsection{University of Hawaii 2.2m Data}

Our multicolor observations of the HD~209458 planetary transit were
made with the University of Hawaii 2.2m telescope on Mauna Kea under
photometric skies. The data were taken with the Tek 2048$^2$ CCD at
$0.22\arcsec$ per pixel.  The brightness of HD~209458, $V = 7.64, \bv
= 0.58$ \citep{hog00}\footnote{For HD 209458 and the comparison stars,
we list \emph{consistent} Johnson $V$ and $\bv$ magnitudes based on
the recommended transformation from the Tycho-2 catalogue $B_T$ and
$V_T$ magnitudes, as described by \citet{esa97}, Vol. 1, Sec. 1.3.},
forced us to defocus the telescope in order not to saturate the
detector. As there were no stars of comparable brightness in the field
of view, we took images of a nearby comparison star before and after
each set of HD~209458 observations. Each observation sequence on
target consisted of five 1~s exposures in each of Johnson~$V$ and
Kron-Cousins~$R$ and $I$, and five 5~s exposures with a $Z$ filter. We
also observed comparison stars that bracketed HD~209458 in airmass,
alternating between HD~210483 ($V = 7.58, \bv = 0.61$) and HD~208156
($V = 8.11, \bv = 0.50$). The comparison star observations were taken
in the same manner as those of the target star.

We performed aperture photometry on the bias-corrected and
flat-fielded CCD images, using a $60\arcsec$ digital aperture. The
magnitudes of the comparison star were plotted against airmass to
define an extinction correction.  We continued observations as far as
airmass 2.1, but a linear extinction correction was sufficient to fit
the data. We also corrected for color-dependent extinction by
interpolating between the extinction corrections derived for the
comparison stars (which straddled HD~209458 in color). 

We converted the derived magnitues into relative fluxes and each set
of five consecutive observations (in each filter) was binned to
produce a single average value, with an uncertainty given by the error
in the mean. The photometry was limited primarily by scintillation
noise, and the achieved relative precision for the time series was
$1.5 \times 10^{-3}$ (1.6 mmag), $9 \times 10^{-4}$ (1.0 mmag), $1.1
\times 10^{-3}$ (1.2 mmag), and $7 \times 10^{-4}$ (0.8 mmag), for the
$V$, $R$, $I$, and $Z$ data, respectively.  For each band, the
off-transit flux level was derived from the statistically-weighted
average of the last three observations, and time series in each color
were normalized to this level to yield the relative fluxes used in the
modeling. Applying the reduction procedure to the comparison star
observations yielded constant light curves as expected.

\subsection{University of Hawaii 0.6m Data}

We obtained Johnson $B$ observations of HD~209458 with the University
of Hawaii 0.6m telescope, in combination with a three channel
photometer, similar to the one described in \citet{klei96}. Using
identical photomultiplier and amplifier-discriminator detectors this
device allows high-speed aperture photometry to be simultaneously
conducted on three selectable regions of the sky, thus enabling
continuous monitoring of a target star, comparison star and sky
background. 

HD~209458 was continuously monitored using 10~s integrations through a
Johnson $B$ filter. A nearby (fainter) star was used for comparison
and guiding, with all channels (including sky) using $30 \arcsec$
apertures. A $B$ filter was not available for the comparison star
channel, so its effective filter was determined by the atmosphere to
the blue and a dichroic beam-splitter to the red (sending the red
light in this channel to a CCD guider), resulting in a bluer and
broader passband than the target channel.  We consequently obtained an
improved atmospheric absorption model for the final data reduction by
observing HD~209458 off-transit on a subsequent night (UT 1999
November 17), during virtually identical photometric conditions over
the same airmass range (1.1 to 1.8) and with the same instrumental
configuration. The HD~209458 sky-corrected instrumental magnitudes
were corrected for color-dependent extinction with a linear airmass
model. These were converted to relative fluxes and binned into
4~minute time bins, with normalization by the weighted average of the
final two data bins, which occured off-transit.  The resulting rms
relative flux residuals were $7 \times 10^{-4}$ (0.8 mmag).

\subsection{STARE Telescope Data}

Additional $R$ observations of HD~209458 were made with the High
Altitude Observatory STARE telescope (Brown \& Kolinski
1999\footnote{Available at
http://www.hao.ucar.edu/public/research/stare/stare.html}; Brown \&
Charbonneau 2000) on the same night, though the target star set at
mid-transit.  The observations and reduction of these data were
identical to those described by \citet{char00}, though the precision
is significantly lower than those observations since the star was
observed at higher airmass.  The data were binned into 5~minute sets,
and converted to relative fluxes by normalizing by the
statistically-weighted average of the off-transit data.  The achieved
relative flux residual rms increased from $3.7 \times 10^{-3}$ (4
mmag) to $6.3 \times 10^{-3}$ (7 mmag) over the course of the night as
the target set.

\begin{figure}
\plotone{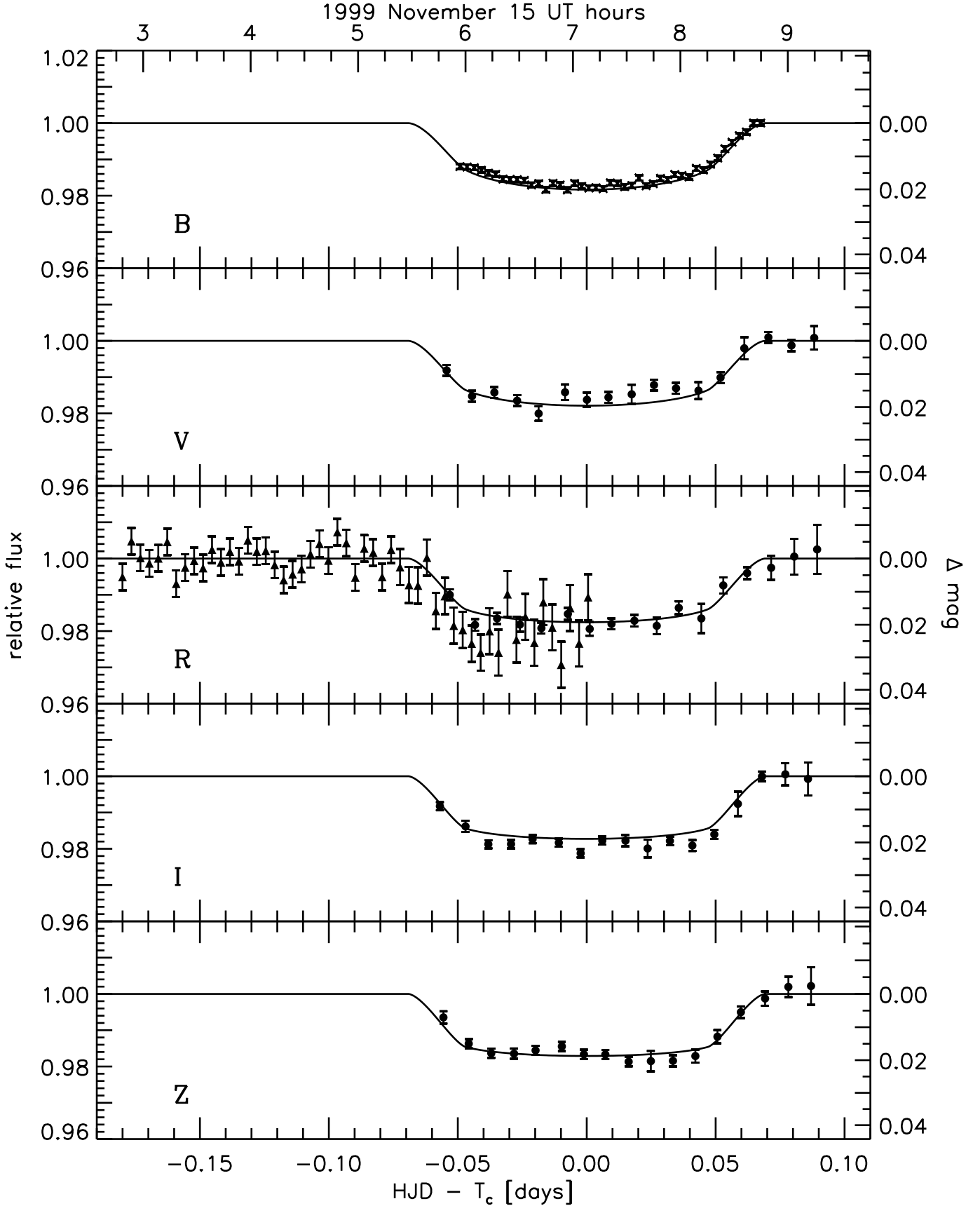}
\caption{Multicolor observations of the HD 209458 planetary transit on
UT 1999 November 15, from the University of Hawaii 2.2m
(\emph{circles}) and 0.6m (\emph{crosses}) telescopes and the High
Altitude Observatory STARE telescope (\emph{triangles}). Uncertainties
in the $B$ data are smaller than the symbols. The error bars do not
include the systematic uncertainties in the zero points, though these
uncertainties were included in the fitting procedure. The solid curves
are our best-fit model for the combined data set. \label{figdata}}
\end{figure}

Our final multicolor photometry is displayed in Figure \ref{figdata}, along
with our best-fit model, described in the next section.

\section{Analysis}

\subsection{Improvement to Orbital Parameters}
Assuming the transit is symmetric, we find the time for the center of
the transit is $T_{c} = 2451497.797 \pm 0.002$~HJD.  The two previous
observations of the full transit \citep{char00} found $T_{c} =
2451430.823 \pm 0.003$~HJD and $T_{c} = 2451437.873 \pm 0.003$~HJD,
which occured, respectively, 19 and 17 orbits earlier than the one we
report here.  From these three measurements of $T_{c}$, we calculate
an orbital period of $P = 3.52495 \pm 0.0003$~days.  We note also that
this period is consistent with the more precise statements of the
period of $P = 3.524736 \pm 0.000045$~days (3.9~s) \citep{cast00} and
$P = 3.524739 \pm 0.000014$~days (1.2~s) \citep{robi00} derived from
\emph{Hipparcos} archive photometry.

\subsection{Analysis of the Multicolor Transit Curve}

Derivation of parameters from transit observations in a single
bandpass is plagued by a degeneracy: the transit curve is described
primarily by its depth and duration, yet it is always possible to fit
these with a larger (smaller) planet provided the stellar radius is
also increased (decreased) and the orbital inclination, $i$, is
decreased (increased). However, by observing the \emph{relative} depth
and shapes of the transit curves in different bands, this degeneracy
can be lifted. Measurement of the relative flux at a given time during
the transit can be combined with a limb-darkening model for the star,
$I_{\lambda}(\mu)$, to derive $\mu$, the cosine of the angle between
the normal to the stellar surface and the line of sight, thus
constraining the inclination. With this constraint, the transit
duration constrains the stellar radius, which when combined with the
transit depth (a measurement of the ratio $R_p / R_s$) constrains the
planetary radius. For example, if one observes that the transit is
deeper in $R$ than in $B$, one would infer that a larger planet passes
close to the limb of a larger star, because the star is relatively
brighter in the red towards the limb.

Because we can observe only the flux integrated over the stellar disk,
we proceed as follows. R. Kurucz (personal communication) generated
model stellar intensities as a function of wavelength and $\mu$.  For
each passband, we multiplied the model intensities by a response
function (see below) and integrated over wavelength to produce
intensity as a function of $\mu$.  For the 2.2m observations, the
response function was the combination of the mirror, CCD, and filter
responses, whereas for the 0.6m and STARE data, we used only the
filter response.  We then produced, for each passband, $C$, a grid of
model light curves, $M_{C}(R_p,R_s,i)$, as a function of the planetary
radius, the stellar radius, and the orbital inclination. The light
curve model follows the description presented in \cite{sack00} and
\cite{char00}, taking into account the finite size of the planet and
the limb-darkening of the star. Since we integrate over the whole
stellar disk and a wide passband, the light curves are relatively
insensitive to small errors in the Kurucz model; we have confirmed
this with fits using parameterized limb-darkening models.

Because all data points of the light curve have been normalized by the
statistically weighted average of the off-transit points, there is a
potential systematic uncertainty in the transit depth, as there are
fewer off-transit points than on-transit points. Thus we introduced a
fourth parameter into the model, a scale factor for the data, $f$,
equivalent to allowing for an unknown magnitude zeropoint.
This nuisance-parameter was subsequently integrated out (see below),
and this integration serves to account properly for the systematic
uncertainty introduced by the paucity of out-of-transit data by
appropriately broadening the likelihood distributions.

We adopt a Bayesian approach for the analysis; in a given passband the
likelihood of a particular model is
\begin{equation}
p_{C}(R_p,R_s,i,f) \propto \exp\left({-\frac{{\chi}^{2}}{2}}\right)
\hat{p}(R_p) \hat{p}(R_s) \hat{p}(i) \hat{p}(f),
\end{equation}
with
\begin{equation}
{\chi}^{2}=\sum_{n=1}^{N}\left(\frac{D_{C}(t_n) - f M_{C}(t_n;R_p,R_s,i)}{\sigma_{C}(t_n)}\right)^{2},
\end{equation}
where $D_{C}(t_n)$ are the observed data with uncertainties
$\sigma_C(t_n)$, and the $\hat{p}$ are the prior probabilities
(priors; see below).  We evaluated $p_{C}$ separately in each passband
for our grid of light curve models spanning the range of
$\{R_p,R_s,i,f\}$ where the likelihood was non-negligible.

As we wish to measure the parameters independently of the previous
results, we assume we have no prior knowledge of $R_{p}$ and $i$.
However, we can estimate $R_{s}$ as follows. With the photometry as
above, the \emph{Hipparcos} parallax of HD~209458 yields $M_{V} = 4.27
\pm 0.11$ \citep{perr97}. Applying a bolometric correction
\citep{bess98} of $-0.06$, appropriate for the stellar type and
metallicity \citep{maze00}, we can calculate the stellar bolometric
flux.  Then from the measured effective temperature of 6000~K
\citep{maze00}, we derive $R_{s} = 1.18 R_{\sun}$.  There is a 10\%
uncertainty in the visual luminosity and a 1\% uncertainty in the
effective temperature, and we estimate a few percent uncertainty in
the bolometric correction. Combining these implies an 8\% uncertainty
in the stellar radius. Thus, in our analysis of the transit curves we
adopt a Gaussian prior for the stellar radius, $\hat{p}(R_s)$
corresponding to $R_s = 1.18 \pm 0.09 R_{\sun}$, and uniform priors
$\hat{p}(R_p)$, $\hat{p}(i)$ and $\hat{p}(f)$, on the other model
parameters. Different choices for the exact shape of the prior on the
stellar radius did not significantly affect the results.

We derived $p_C$ independently for each of the six passbands: the UH
0.6m $B$ data; the UH 2.2m $V$, $R$, $I$ and $Z$ data; and the STARE
$R$ data. The minimum reduced $\chi^2$ is very near unity for four of
the six passbands, $B: \chi^2_\nu = 0.97$ (with 39 degrees of
freedom), $V: \chi^2_\nu = 1.16$ (13), STARE $R: \chi^2_\nu = 1.03$
(49), $I: \chi^2_\nu = 1.04$ (13).  The two exceptions are the UH 2.2m
$R$ and $Z$ data, which have minimum $\chi^2_\nu$ of 2.85 and 2.97,
respectively for 13 degrees of freedom each. As these are the two
passbands with the smallest statistical uncertainty, we adopt the
plausible hypothesis that the poor fits are caused by an unidentified
systematic uncertainty (at the level of $\sim 1$ mmag), rather than a
failing of the light curve models. Data of higher precision will
resolve this issue, but for the time being we have scaled the
uncertainties for the UH 2.2m $R$ and $Z$ data to give a minimum
reduced $\chi^2$ of unity, thus broadening the derived likelihood
distributions.

To derive combined constraints from the individual likelihoods, we
first integrate each $p_C$ over the nuisance parameter $f$, leaving
correctly broadened three-dimensional likelihood distributions over
$\{R_p, R_s, i\}$. Multiplying these distributions together (but with
the priors included only once) yields our combined likelihood; the
peak of this distribution is our best-fit model and the light curves
corresponding to this model are shown as the solid lines in Figure
\ref{figdata}. We note that the uncertainties in our derived
parameters are highly correlated and to underscore this important
fact, we present contours of the two-dimensional likelihood
distributions (integrated over the third variable) in Figure
\ref{figlike}. To correctly account for these correlations in
determining the individual likelihoods for each of the parameters
separately, we integrate the three-dimensional distribution
over the other two parameters. These one-dimensional distributions are
well-fit by Gaussians, with $R_p = 1.55 \pm 0.10 \, R_{\rm Jup}$, $R_s
= 1.27 \pm 0.05 \, R_{\sun}$, and $i = 85.9 \pm 0.5 \degr$. 

\section{Discussion and Conclusion}

Our derived parameters are consistent with the results of
\cite{maze00}, with our multicolor data roughly halving the
uncertainties presented there, and favoring a slightly larger planet
and star. Combined with the orbital data of \cite{maze00}, our results
imply a mean planetary density of $\rho = 0.23 \pm 0.05 \; {\rm g \,
cm^{-3}}$.

We are also in a position to make a comparison between the
observations and models for HD~209458b. \cite{burr00} present
evolutionary tracks of radii versus time for close-in extrasolar giant
planets, taking into account the effects of high stellar insolation in
retarding the planet's contraction. Previous measurements of the
radius of HD~209458b were not precise enough to differentiate between
models with low and moderate albedos (models A and B of \cite{burr00}
in their Figure 1, with Bond albedos of 0.0 and 0.5,
respectively). Our more precise measurement favors a larger planet,
and is in very good agreement with model A, which has a radius close
to 1.5 $R_{\rm Jup}$ (at the approximate age of the system of several
Gyr). Our derived radius disfavors model B, for which the radius is
predicted to be near 1.3 $R_{\rm Jup}$. This result provides evidence
for a low albedo for HD~209458b.

\begin{figure}
\epsscale{0.55}
\plotone{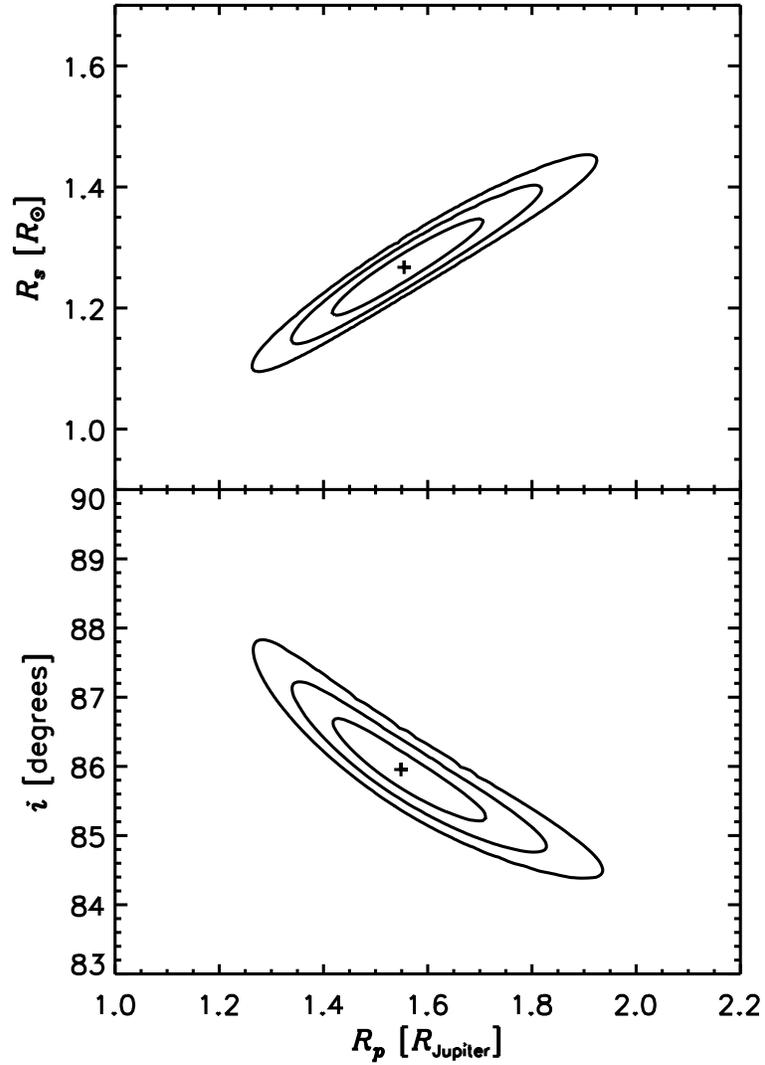}
\caption{Contour plots of the marginalized
two-dimensional likelihood distributions, illustrating that
all three parameters are highly correlated. The peak position is shown
by a cross, and the contours displayed correspond to 68.3, 95.4 and
99.7\% confidence levels. \label{figlike}}
\end{figure}

\begin{figure}
\epsscale{0.55}
\plotone{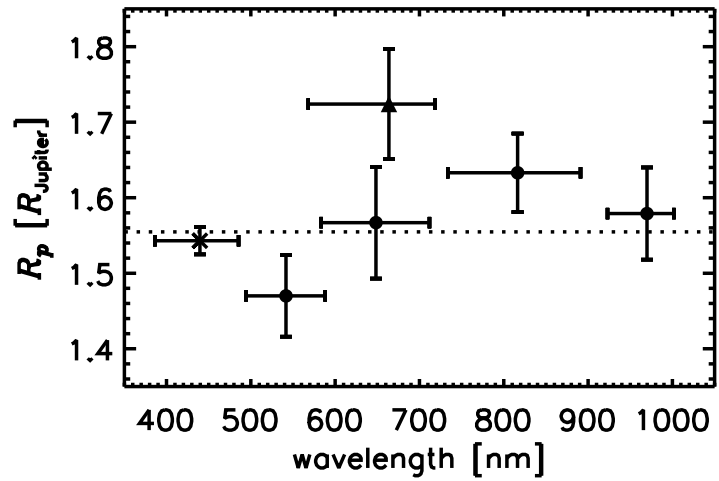}
\caption{Planetary radii derived independently for each
of the six data sets, with the stellar radius and orbital inclination
\emph{fixed} at $R_s = 1.27 R_{\rm Jup}$ and $i = 85.9\degr$,
respectively. The symbols are as in Figure \ref{figdata}, with the
points placed at the effective wavelength of each bandpass and the
horizontal error bars indicating where the transmission falls to half
the peak value. The vertical error bars do not include the uncertainty
common to all the points from the uncertainties in the stellar radius
and orbital inclination. The dotted line shows the the best-fit
planetary radius with all the data combined. \label{figrad}}
\end{figure}

It is also of interest to see if we can detect any differences in the
planetary radius as a function of color.  The value of $R_{p}$ as
derived from transit observations may vary with the passband, since
the planet would appear slightly larger when observed at wavelengths
where the atmosphere contains strong opacity sources
\citep{brow00,burr00,seag00}. To explore this possibility, in Figure
\ref{figrad} we present the peak positions of the one-dimensional
conditional likelihood distributions for the planetary radius in each
passband, with the stellar radius and inclination \emph{fixed} at
their best-fit values. The uncertainties shown then do not incorporate
the correlations exhibited in Figure \ref{figlike}, but they are
correct relative to each other. For instance, if the stellar radius
were increased, \emph{all} of the points on Figure \ref{figrad} would
increase together. Even so, the data are not precise enough to show
evidence for any variation in the planetary radius with
wavelength. Future observations should be able to improve the
measurement precision, and provide a tool for learning about the
planetary atmosphere.

The discovery of an extrasolar planet transiting its host star has
heralded a new era in a young field. While precision radial velocities
reveal much about orbital characteristics, the addition of transit
observations yield understanding about the physical characteristics of
the planets themselves. We eagerly anticipate further exploration of
extrasolar planets with continued observations of this shadowy nature.

\acknowledgements 

We thank R. Kurucz for producing the stellar model, R. Noyes and
T. Mazeh for valuable discussions, J. Dvorak for his skill operating
the UH 2.2m, and the anonymous referee for useful
suggestions. D.~J.~S. and T.~S. were observing at the UH 0.6m as part
of a Whole Earth Telescope campaign and acknowledge financial support
from IITAP at Iowa State University, partially funded by
UNESCO. D.~C. is supported in part by a Newkirk Fellowship of the High
Altitude Observatory.  This work was supported in part by NASA grant
W-19560. S.~J. is supported through an NSF Graduate Research
Fellowship.

%
%

\begin{thebibliography}{}
\bibitem[Bessell, Castelli, \& Plez(1998)]{bess98}Bessell, M.~S., Castelli, F.,
\& Plez, B. 1998, \aap, 333, 231
\bibitem[Brown \& Charbonneau(2000)]{brow00}Brown, T.~M., \&
Charbonneau, D. 2000, in ASP Conf. Ser., Disks, Planetesimals,
and Planets, eds. F. Garz\'on and T. J. Mahoney (San Francisco: ASP) 
(astro-ph/0005009)
\bibitem[Burrows et al.(2000)]{burr00}Burrows, A., Guillot, T., 
Hubbard, W.~B., Marley, M.~S., Saumon, D., Lunine, J.~I., 
\& Sudarsky, D. 2000, \apjl, in press (astro-ph/0003185)
\bibitem[Castellano et al.(2000)]{cast00}Castellano, T., 
Jenkins, J., Trilling, D.~E., Doyle, L., \& Koch, D. 2000,
\apjl, 532, L51
\bibitem[Charbonneau et al.(2000)]{char00}Charbonneau, D., Brown, T. M.,
Latham, D. W., and Mayor, M. 2000, \apjl, 529, L45
\bibitem[ESA(1997)]{esa97}European Space Agency 1997, The Hipparcos and Tycho
Catalogues (Paris: ESA), SP-1200 
\bibitem[Henry et al.(2000)]{henr00}Henry, G. W., Marcy, G. W., Butler, R. P.,
and Vogt, S. S. 2000, \apjl, 529, L41
\bibitem[H{\o}g et al.(2000)]{hog00}H{\o}g, E., Fabricius, C., Makarov, V. V.,
Urban, S., Corbin, T., Wycoff, G., Bastian, U., Schwekendiek, P., \& Wicenec,
A. 2000, \aap, 355, L27
\bibitem[Kleinman, Nather, \& Phillips(1996)]{klei96}Kleinman, S. J., Nather,
R. E., and Phillips, T. 1996, \pasp, 108, 356
\bibitem[Mazeh et al.(2000)]{maze00}Mazeh, T., et al. 2000, \apj, 532,
55
\bibitem[Perryman et al.(1997)]{perr97}Perryman, M. A. C., et
al. 1997, \aap, 323, L49
\bibitem[Robichon \& Arenou(2000)]{robi00}Robichon, N., \& Arenou, F. 2000,
\aap, 355, 295
\bibitem[Sackett(1999)]{sack00}Sackett, P. D. 1999, in NATO/ASI Ser., 
Planets Outside the Solar System: Theory and Observations, 
ed. J.-M. Mariotti \& D. Alloin (Dordrecht: Kluwer), 189
\bibitem[Seager \& Sasselov(2000)]{seag00}Seager, S., \& Sasselov,
D. D. 2000, \apj, submitted (astro-ph/9912241)
\end{thebibliography}
\end{document}